\documentclass[%
 reprint,
superscriptaddress,
 amsmath,amssymb,
 aps,
 prl,
]{revtex4-2}

\usepackage{graphicx}
\usepackage{dcolumn}
\usepackage{bm}
\usepackage{physics}
\usepackage{booktabs}
\usepackage{siunitx}
\usepackage{titlesec}

\begin{document}

\title{A faithful solid-state spin-wave quantum memory for polarization qubits}

\author{Ming Jin}

\author{You-Zhi Ma}

\author{Zong-Quan Zhou}
 \email{email: zq\_zhou@ustc.edu.cn; cfli@ustc.edu.cn}
\author{Chuan-Feng Li}
 \email{email: zq\_zhou@ustc.edu.cn; cfli@ustc.edu.cn}
\author{Guang-Can Guo}
\affiliation{
 CAS Key Laboratory of Quantum Information,\\University of Science and Technology of China, Hefei, 230026, P. R. China
}
\affiliation{
 CAS Center for Excellence in Quantum Information and Quantum Physics, University of Science and Technology of China, Hefei, 230026, P. R. China
}

\maketitle
Memory based quantum communication can overcome the distance limit, by adopting either the approach of a quantum repeater \cite{RevModPhys.83.33} or a transportable quantum memory \cite{6hour,ma2021one}. Photonic quantum memories have been demonstrated with various physical systems, including single atoms inside optical cavities \cite{specht2011single}, atomic gases \cite{yu2020entanglement} and impurities in solids \cite{Efficient_2010}.
As an emerging solid-state platform, rare-earth-ion doped solids have enabled optical quantum storage with remarkable performances \cite{Efficient_2010,liu2021heralded,saglamyurek2011broadband,zhong2017nanophotonic}. 
Moreover, the rare-earth-ion doped crystal, i.e., Eu$^{3+}$:Y$_2$SiO$_5$, is currently the only candidate for a transportable photonic quantum memory, due to its hours-long optical storage times \cite{ma2021one}. Despite these remarkable achievements, rare-earth-ion doped crystals typically have birefringence and anisotropic absorption which make the storage of polarization-qubits become a great challenge. Pioneering works \cite{zhou2012,Gisin2012,Riedmatten2012,Pr_2015} have demonstrated polarization qubit storage based on the two-level atomic frequency comb (AFC) with a predetermined storage time. The only exception is the demonstration of the spin-wave AFC protocol in a Eu$^{3+}$:Y$_2$SiO$_5$ crystal in a double-pass configuration \cite{NJP_2015}. Unfortunately, the achieved storage process fidelity ($\sim$ 76.2\%) cannot surpass the maximum fidelity achievable with a classical measure-and-prepare strategy \cite{Riedmatten2012}. What's more, the double-pass configuration could introduce unwanted losses, instabilities and complexities for practical applications.
Here we demonstrate a spin-wave solid-state quantum memory for photonic polarization qubits using a single piece of crystal in a single-pass configuration which has a storage process fidelity of 91.9(24)\% for single-photon level weak coherent pulses. 

Our approach requires a single crystal which provides a near-uniform absorption for two orthogonal polarization states. Eu$^{3+}$ ions at site 2 of Y$_2$SiO$_5$ crystals can fulfill this requirement \cite{Eu_2003} and the price to pay is the significantly reduced sample absorption as compared to that of site-1 Eu$^{3+}$ ions. As a result, the widely employed AFC protocol is challenging to be implemented here since the efficiency would be low due to the reduced absorption caused by the spectral tailoring process. Our recently-proposed spin-wave storage protocol, noiseless photon echo (NLPE), can be useful in such circumstance since it requires no spectral tailoring and makes a complete use of the original sample absorption \cite{NLPE_2021}.

\begin{figure*}
\centering
\includegraphics[width=\textwidth]{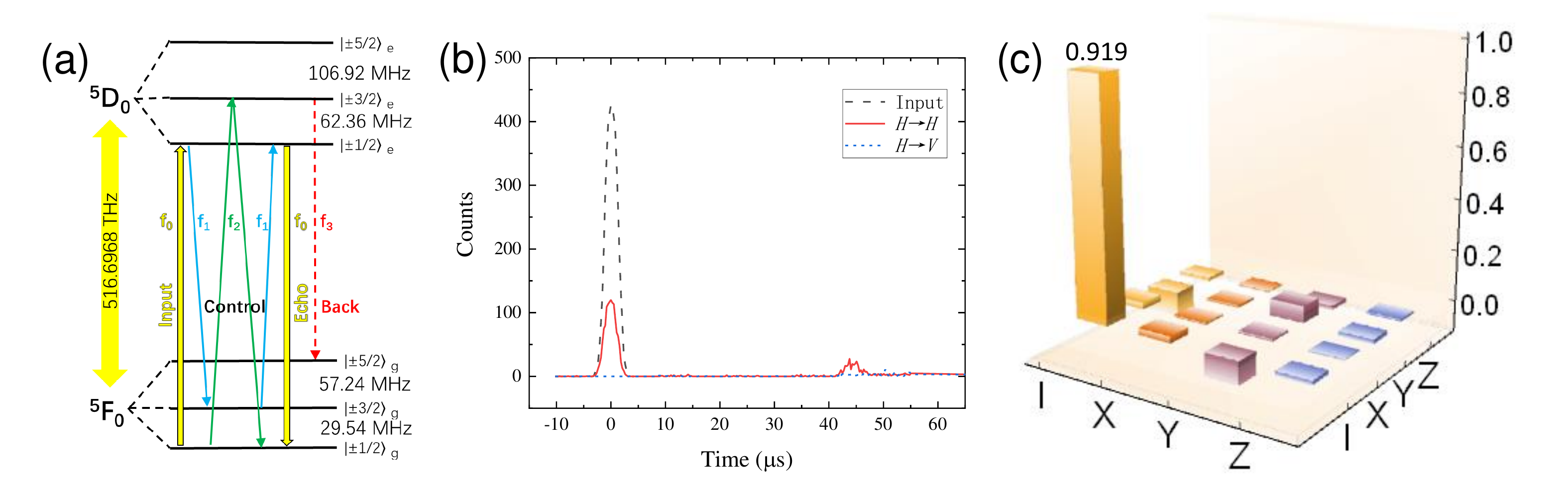}
\caption{(Color online) (a) Energy level structure of $^{151}$Eu$^{3+}$ at site 2 of Y$_2$SiO$_5$ crystals at zero magnetic field. The laser is resonant with the $^{7}$F$_{0}$ $\leftrightarrow$ $^{5}$D$_{0}$ transition with a frequency of 516.6968 THz. The solid arrows (yellow, blue and green) indicate the transitions occurred during the NLPE storage sequence \cite{NLPE_2021}. The dashed arrow (red) represents the back pump light ( f$_3$) used in the initialization process. (b) Photon counting histograms with the input state of $\ket{H}$ and projection measurements in $\ket{H}$ (red solid line) and $\ket{V}$ (blue dotted line). The black dashed line represents the input pulse. (c) The real part of the reconstructed process matrix. $\chi_{00}$ is 0.919 and all imaginary numbers of $\chi$ are close to zero with the maximum of 0.055.}
\label{fig:hutu}
\end{figure*}

To implement spin-wave quantum storage, a complete knowledge about the hyperfine energy levels is required. Here, spectral-hole burning techniques are employed to characterize the $^{7}$F$_{0} \longrightarrow ^{5}$D$_{0}$ transition of $^{151}$Eu$^{3+}$ ions at site 2 of Y$_2$SiO$_5$ crystals (Fig.~\ref{fig:hutu}a) and details are provided in the Supplementary Materials. Based on this level structure, a single class of ions can be isolated through optical pumping and the relative transition strengths can be determined after polarization of the hyperfine state (Tab.~\ref{BR}).

The experimental setup is similar to that in our recent work \cite{NLPE_2021}. The only exception is the preparation and detection of polarization states, which are implemented by standard wave plates and polarization beam splitters \cite{zhou2012}. An 0.1\%-doped $^{151}$Eu$^{3+}$:Y$_2$SiO$_5$ crystal is employed as the memory crystal (MC) which is cut along D2 axis with a length of 10 mm. To avoid the associated hole at the frequency of one control pulse in NLPE, we use a 20-mm $^{153}$Eu$^{3+}$: Y$_2$SiO$_5$ crystal with a doping level of 0.2\% as the filter crystal (FC). FC is double passed to provide an effective absorption depth of 5.2.

\begin{table}[b]
\centering
\caption{\bf Relative transition probabilities}
\begin{tabular}{cccc}
\hline
 &$\ket{\pm 1/2}_{\rm e}$&$\ket{\pm 3/2}_{\rm e}$&$\ket{\pm 5/2}_{\rm e}$ \\
\hline
$\ket{\pm 1/2}_{\rm g}$& 0.771 & 0.144 & 0.085 \\
$\ket{\pm 3/2}_{\rm g}$& 0.184 & 0.727 & 0.089 \\
$\ket{\pm 5/2}_{\rm g}$& 0.045 & 0.129 & 0.826 \\
\hline
\end{tabular}
  \label{BR}
\end{table}

According to Tab.~\ref{BR}, the NLPE protocol \cite{NLPE_2021} is implemented with a four-level system presented in Fig.~\ref{fig:hutu}. The input pulse with the center frequency of f$_0$ is absorbed by MC and the echo is retrieved after the application of two pairs of control pulses with the center frequencies of f$_1$ and f$_2$. After the spectral initialization process which isolates a well-defined four-level system inside the inhomogeneously broadened absorption line, the absorption peak for the input has a full width at half maximum of 800 kHz and the absorption depth is 1.32 for $\ket{H}$ and 1.53 for $\ket{V}$, respectively. Here $\ket{H}$ ($\ket{V}$) denotes the polarization state that is parallel to the D1 axis (b axis). The polarization state of the pump is chosen as $\ket{V}$. During the initialization process, the pump light for FC burns a single hole with a bandwidth of 1 MHz which is resonant with the signal. More experimental details are provided in the Supplementary Materials.

Polarization qubits are encoded with weak coherent pulses with an average photon number of $1.10\pm0.04$ photons per input pulse. Four input states $\ket{H}$, $\ket{V}$, $\ket{D}=(\ket{H}+\ket{V})/\sqrt{2}$ and $\ket{R}=(\ket{H}+i\ket{V}/\sqrt{2})$ are stored and the output is projected to these four states. An example of the photon-counting histogram is shown in Fig.~\ref{fig:hutu}b. Given a detection window of 1.57 $\mu$s, the measured storage efficiency is $4.81\pm0.46\%$ for $\ket{H}$ and $5.31\pm0.38\%$ for $\ket{V}$, with the storage time of 44.5 $\mu$s. 
The average noise floor is measured (without input pulses) as 0.003 photons per trial and the signal-to-noise ratio (SNR) is $9.3\pm1.6$ for $\ket{H}$ and $10.2\pm1.7$ for $\ket{V}$. The noise mainly comes from the spontaneous decay from the excited state $\ket{\pm 3/2}_{\rm e}$ to the ground state $\ket{\pm 3/2}_{\rm g}$ during the storage in the optically excited state \cite{NLPE_2021}. 
To evaluate the qubit storage performance in a rigorous way, we perform the quantum process tomography of this device \cite{zhou2012,NJP_2015}. The process matrix $\chi$ is reconstructed using the maximum likelihood estimation \cite{zhou2012,NJP_2015} and the real part of $\chi$ is shown in Fig.~\ref{fig:hutu}c. The storage fidelity is 91.9$\%$ with one standard deviation of 2.4$\%$ estimated by the Monte Carlo simulation \cite{zhou2012,NJP_2015}. 
Our measured fidelity significantly outperforms the previous result (76.2\%) reported in Ref. \cite{NJP_2015} albeit with a shorter storage time, and is well above the strict classical bound 84.2$\%$ \cite{Riedmatten2012,NLPE_2021}. The measured fidelity is primarily limited by the SNR while the reduced fidelity caused by the residual anisotropic absorption is less than 0.1\% according to our simulation. To further eliminate any impact from the residual anisotropic absorption, we can operate our memory with a sample absorption depth near to 2 where the storage efficiency is not sensitive to the change of absorption \cite{NLPE_2021}. While completely suppressing the noise is not possible in zero fields, operating at a high field can in principle eliminate the remaining noise by employing appropriate selection rules to forbid the unwanted optical decay path \cite{NLPE_2021}. One shortage of the NLPE memory is the temporal multimode capacity as compared to the AFC memory \cite{NJP_2015}. This problem can be overcome by the spatial multiplexing based on fabricating multiple storage channels using fs-laser for example \cite{PhysRevLett.125.260504}. Our compact design of polarization qubit memory is particularly favorable in fabrication of integrated devices, as compared to the spatially-splitted storage approach \cite{zhou2012,Gisin2012,Riedmatten2012,liu2021heralded} or the double-passed approach \cite{NJP_2015}.

In summary, spin-wave storage of polarization qubits is demonstrated with a single piece of  $^{151}$Eu$^{3+}$:Y$_2$SiO$_5$ crystal in a single-pass configuration. The memory protocol and the strategy of polarization storage developed here are universal for atomic ensemble systems. Due to the compact design and the built-in spin-wave storage, the current approach could be further extended to storage times of hours \cite{ma2021one} by employing zero-first-order-Zeeman (ZEFOZ) magnetic fields and dynamical decoupling (DD) to protect the spin coherence. In that case, we need to further identify proper ZEFOZ fields where the sample provides near-uniform absorption. When operating at ZEFOZ fields, the sample absorption is severely limited because that magnetic level degeneracy has been removed and the sample size is limited to ensure a high homogeneity for both the DC field and the RF fields from DD pulses \cite{6hour,ma2021one}. An efficient 4-level system need to be identified to implement the NLPE protocol which shows significant advantages in efficiencies with weak-absorbing sample. This compact device for polarization qubit storage could provide a useful solution for the construction of a long-lived transportable quantum memory \cite{6hour,ma2021one} and the memory-based quantum networks \cite{RevModPhys.83.33,liu2021heralded,saglamyurek2011broadband}.

\section*{Conflict of interest}
The authors declare that they have no conflict of interest.

\section*{Acknowledgement}

This work is supported by the National Key R\&D Program of China (No. 2017YFA0304100), the  National Natural Science Foundation of China (Nos. 11774331, 11774335, 11821404 and 11654002) and the Fundamental Research Funds for the Central Universities (No. WK2470000026 and No. WK2470000029). Z.-Q.Z acknowledges the support from the Youth Innovation Promotion Association CAS.

\section*{Author contributions}
Zong-Quan Zhou designed the experiment; Ming Jin performed the experiment and analyzed the data with the help from You-Zhi Ma. Ming Jin and Zong-Quan Zhou wrote the manuscript. Zong-Quan Zhou and Chuan-Feng Li supervised the project. All authors discussed the experimental procedures and results.

\section*{Appendix: Supplementary materials}

\renewcommand{\figurename}{Supplementary Fig.}
\renewcommand{\thefigure}{\arabic{figure}}

\subsection*{Supplementary Note 1 - \\Determination of the order of hyperfine levels}

Before determining the order of the hyperfine levels, we 
obtain the excited-state (ground-state) hyperfine splittings by the analysis of the antiholes (holes) based on the spectrum of single-hole burning \cite{Spectroscopic_2012}. The ground-state hyperfine splittings are 29.54 MHz and 57.24 MHz, and the excited-state hyperfine splittings are 62.36 MHz and 106.92 MHz.

In order to determine the order of the hyperfine levels, we perform spectral-hole burning measurements \cite{Spectroscopic_2012,Hyperfine_2004} based on a combination of single-frequency burning and dual-frequency burning techniques. 
By comparing the size of the antiholes at certain positions, we can obtain the order of the hyperfine levels both in the ground state and the excited state. This method is simple and intuitive, without any complex analysis.

First, we determine the order of hyperfine levels in the ground state. we use the pump beams at two frequencies simultaneously to burn the spectral holes in the sample. The spacing of two frequencies is one of the hyperfine splittings in the ground state (29.54 MHz, 57.24 MHz or 86.78 MHz). Two different ground states couple to the same excited state. Due to the resonance with two pump beams, some classes of ions have two hyperfine ground states that are excited continuously. Therefore, these ions are pumped into the third hyperfine ground state and the antiholes related to this state are enhanced.

\begin{figure}
\includegraphics [width= 0.9 \columnwidth]{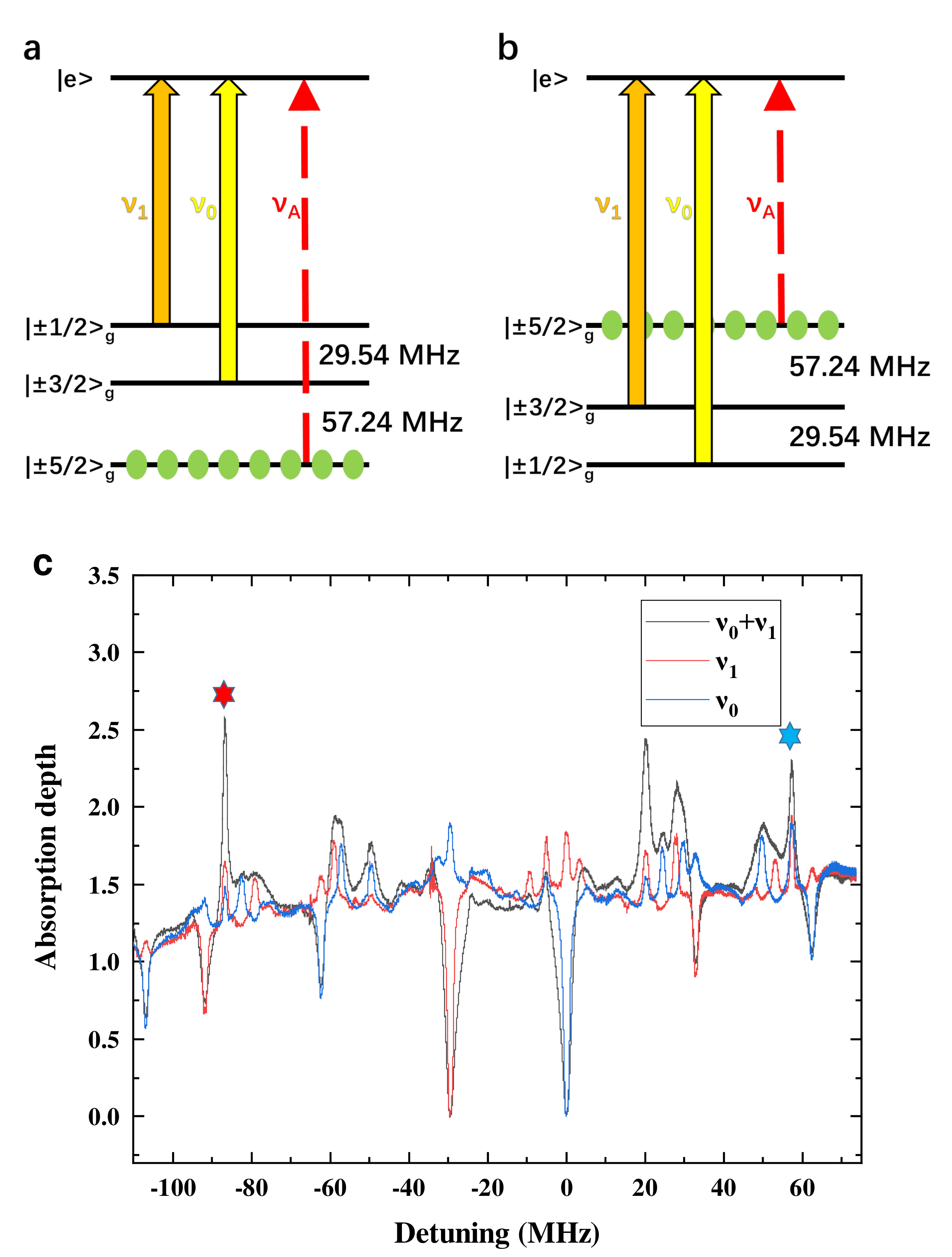}
\caption{\label{fig:jitainew} \textbf{a} A possible order of ground states with the 29.54-MHz splitting above the 57.24-MHz splitting. The resonance with the pump beams at the frequencies $\nu_0$ and $\nu_1$ generates an enhanced antihole at the frequency $\nu_\mathrm{A}$ = $\nu_0+57.24$ MHz. \textbf{b} The other possible order of ground states. The resonance with the light at $\nu_0$ and $\nu_1$ generates an enhanced antihole at the frequency $\nu_\mathrm{A}$ = $\nu_0-86.78$ MHz. \textbf{c} The hole burning spectra using the single-frequency pump beam at $\nu_0$ (blue), single-frequency pump beam at $\nu_1$ (red), and dual-frequency pump at $\nu_0$ and $\nu_1$ (black). Zero detuning in the figure corresponds to the frequency $\nu_0$. The antihole at 57.24 MHz is marked with the blue star and the antihole at $-86.78$ MHz is marked with the red star.}
\end{figure}

We use the pump beams at the frequencies $\nu_0$ and $\nu_1$ = $\nu_0-29.54$ MHz. As shown in Supplementary Fig.~\ref{fig:jitainew}a and \ref{fig:jitainew}b, the enhanced antiholes will have different frequencies assuming the different order of the hyperfine levels in the ground state. In both cases, the population in the ground states $\ket{\pm 1/2}_{\rm g}$ and $\ket{\pm 3/2}_{\rm g}$ are pump into the ground state $\ket{\pm 5/2}_{\rm g}$. The antihole at the frequency $\nu_\mathrm{A}$ should be enhanced, which is corresponding to the transition from $\ket{\pm 5/2}_{\rm g}$ to the same excited state. Due to the contribution of three classes of ions to the antiholes $\ket{\pm 5/2}_{\rm g}$ $\leftrightarrow$ $\ket{\pm i/2}_{\rm e}$ (i = 1, 3, 5), one can clearly see the enhanced antiholes despite the relative strengths of these transitions.

\begin{figure}
\includegraphics [width= 0.9 \columnwidth]{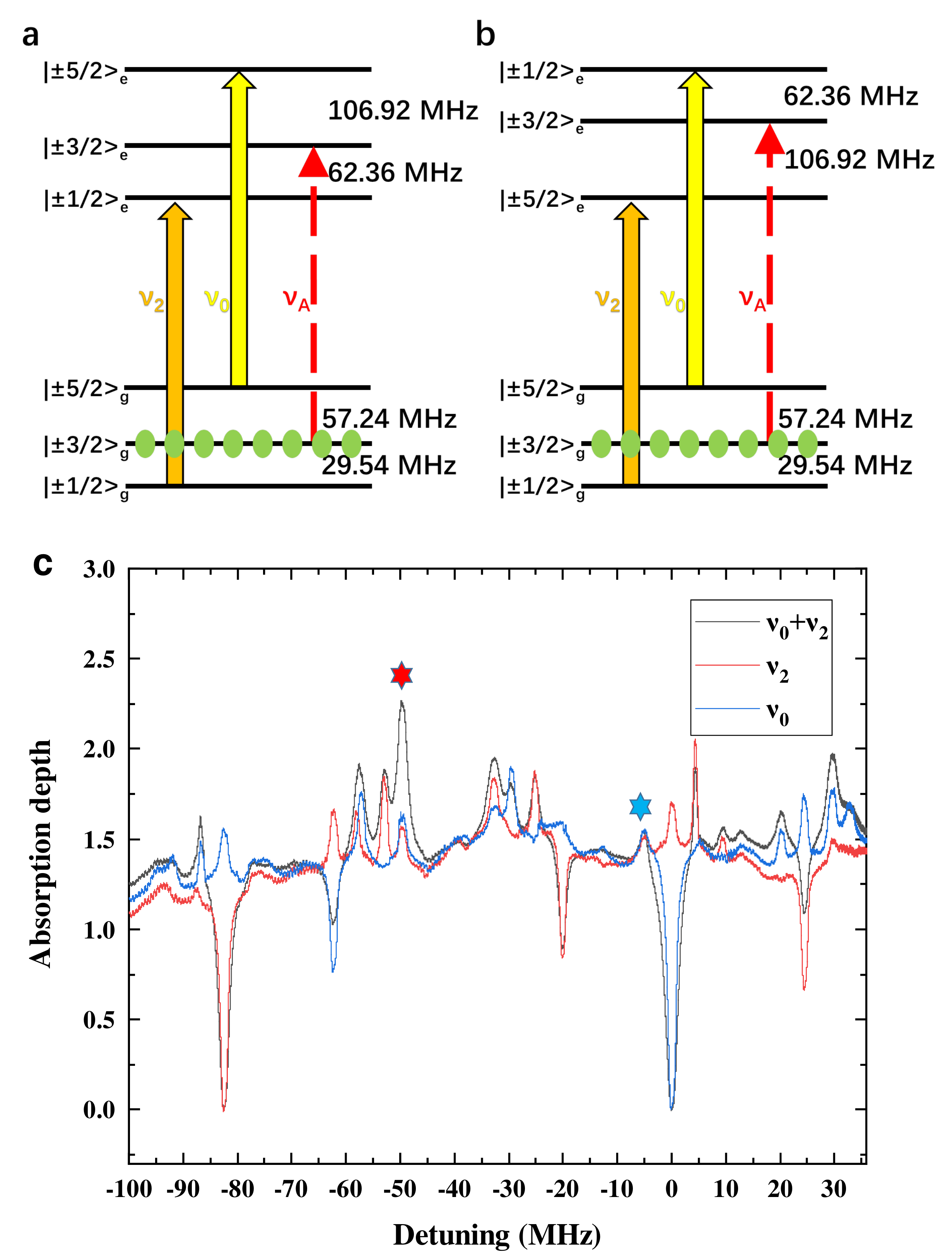}
\caption{\label{fig:jifatainew} \textbf{a} A possible order of excited state with the 106.92-MHz splitting above the 62.36-MHz splitting. The resonance with the pump beams at the frequencies $\nu_0$ and $\nu_2$ generates an enhanced antihole at the frequency $\nu_\mathrm{A}$ = $\nu_0-49.68$ MHz. \textbf{b} The other possible order of excited state. The resonance with the light at $\nu_0$ and $\nu_2$ generates an enhanced antihole at the frequency $\nu_\mathrm{A}$ = $\nu_0-5.12$ MHz. \textbf{c} The hole burning spectra using the single-frequency burning at $\nu_0$ (blue), single-frequency burning at $\nu_2$ beam (red), and dual-frequency buring at $\nu_0$ and $\nu_2$ (black). Zero detuning corresponds to the frequency $\nu_0$. The antihole at $-5.12$ MHz is marked with the blue star and the antihole at $-49.68$ MHz is marked with the red star.}
\end{figure}

The hole burning spectrum using $\nu_0$ and $\nu_1$ is shown in Supplementary Fig.~\ref{fig:jitainew}c. Unfortunately, there are  enhanced antiholes at the frequencies of both $\nu_0+57.24$ MHz and $\nu_0-86.78$ MHz. A comparison with the single-frequency burning spectra is required to obtain the ordering information. The absorption depth of an antihole is d$_0$ + $\Delta$d, where d$_0$ is the original absorption depth and $\Delta$d is the increased absorption depth caused by the spectral hole burning. If there are no ions that are simultaneously resonance with two frequencies and contribute to the antihole, $\Delta$d of the antihole with dual-frequency burning is simply the sum of the increased absorption depth caused by the single-frequency burning, i.e., $\Delta$d = $\Delta$d$_0$ + $\Delta$d$_1$, where $\Delta$d$_0$ ($\Delta$d$_1$) is the increased absorption depth caused by the single-frequency burning at $\nu_0$ ($\nu_1$). For simplicity, we assume that there is only one excited hyperfine state $\ket{e}$. $\Delta$d at $\nu_\mathrm{A}$ is proportional to the product of the excess population $\Delta$n in the ground state $\ket{\pm 5/2}_{\rm g}$ and the transition probability $\gamma_{5e}$, i.e., $\Delta$d $\propto$ $\Delta$n $\cdot$ $\gamma_{5e}$ \cite{Spectroscopic_2012}, where $\gamma_{ie}$ (i = 1, 3, 5) denotes the transition probability of the $\ket{\pm i/2}_{\rm g}$ $\leftrightarrow$ $\ket{e}$ transition. 

We assume the population initially in the three ground states is the same, i.e. N/3, where N is the total number of ions. For single-frequency hole burning at $\nu_0$ ($\nu_1$), the population in the ground state $\ket{\pm 3/2}_{\rm g}$ ($\ket{\pm 1/2}_{\rm g}$) is all pumped away and distributed in the ground states $\ket{\pm 1/2}_{\rm g}$ ($\ket{\pm 3/2}_{\rm g}$) and $\ket{\pm 5/2}_{\rm g}$ according to the transition probabilities. Then the excess population $\Delta$n$_0$ ($\Delta$n$_1$) caused by the single-frequency hole burning at $\nu_0$ ($\nu_1$) is given by:
\begin{equation}
\Delta n_0 = \frac{1}{3}N\frac{\gamma_{5e}}{\gamma_{1e}+\gamma_{5e}}, 
\Delta n_1 = \frac{1}{3}N\frac{\gamma_{5e}}{\gamma_{3e}+\gamma_{5e}}.
\end{equation} 
For the excess population $\Delta$n caused by the dual-frequency burning at $\nu_0$ and $\nu_1$, the population in the ground state $\ket{\pm 1/2}_{\rm g}$ and $\ket{\pm 3/2}_{\rm g}$ are all pumped into $\ket{\pm 5/2}_{\rm g}$ and $\Delta$n is 2N/3. Obviously, $\Delta$n $>$ $\Delta$n$_0$ + $\Delta$n$_1$. As a result, $\Delta$d $>$ $\Delta$d$_0$ + $\Delta$d$_1$ is expected. According to the data presented in Supplementary Fig.~\ref{fig:jitainew}c, the antihole at $\nu_0-86.78$ MHz satisfies our expectation, indicating that the 57.24-MHz splitting is above the 29.54-MHz splitting in the ground state.

Based on the determined ground-state order, we further perform the single-frequency burning and dual-frequency burning experiments to determine the order of hyperfine level in the excited state. The two frequencies are in resonance with two different ground states coupling to different excited states. We use the dual-frequency pump at the frequencies $\nu_0$ and $\nu_2$ =  $\nu_0-82.5$ MHz. As shown in Supplementary Fig.~\ref{fig:jifatainew}a and \ref{fig:jifatainew}b, we can determine the order by the frequency of the enhanced antihole. Similar to the analysis presented above, we can also infer $\Delta$d $>$ $\Delta$d$_0$ + $\Delta$d$_2$, where $\Delta$d is the increased absorption depth caused by the dual-frequency burning at $\nu_0$ and $\nu_2$, $\Delta$d$_0$ ($\Delta$d$_2$) is the increased absorption depth caused by the single-frequency burning at $\nu_0$ ($\nu_2$). The result is shown in Supplementary Fig.~\ref{fig:jifatainew}c and the antihole at $\nu_0-49.68$ MHz satisfies the expectation. As a result, the level structure shown in Supplementary Fig.~\ref{fig:jifatainew}a is the correct one.

\subsection*{Supplementary Note 2 - \\Measurements on the relative transition strengths}

Following the method developed in \cite{Spectroscopic_2012}, a single class of ions is isolated using pump light with three frequencies. We choose three transitions where two ground-state levels couple to the same excited-state level and the third ground-state level couples to another different excited-state level. The first step is so-called class cleaning, where pump light centering at $\omega_0$, $\omega_1$ and $\omega_2$ simultaneous incident on the sample to select a single class of ions. The second step is so-called spin polarization where pump light centering at two frequencies can polarize all ions to a certain ground-state level. The final step is the narrow-hole burning, 
where a narrow hole is burned at the polarized ground-state level and those ions related the narrow hole will be polarized to another chosen ground-state level. Comparing the spectrum after the spin polarization and the final spectrum, we can see the enhanced absorption from the chosen ground-state level to three excited-state levels.

For example, as shown in Supplementary Fig.~$\ref{fig:BRnew}$a, we can use pump light centering at $\omega_0$, $\omega_1$ = $\omega_0$ +57.24 MHz and $\omega_2$ = $\omega_0-20.14$ MHz, which correspond to the transitions $\ket{\pm 5/2}_{\rm g}$ $\leftrightarrow$ $\ket{\pm 5/2}_{\rm e}$, $\ket{\pm 3/2}_{\rm g}$ $\leftrightarrow$ $\ket{\pm 5/2}_{\rm e}$ and $\ket{\pm 1/2}_{\rm g}$ $\leftrightarrow$ $\ket{\pm 3/2}_{\rm e}$, respectively. In class cleaning and spin polarization procedures, the light are swept over 4 MHz and a single class of ions is polarized to $\ket{\pm 1/2}_{\rm g}$ ground state with 4-MHz bandwidth. In the third step, we burn a narrow hole at $\omega_2$ and sweep the pump light centering at $\omega_0$. These chosen ions will be polarized to $\ket{\pm 3/2}_{\rm g}$ ground state and the enhanced antiholes related to this ground state can be observed. As shown in Supplementary Fig.~$\ref{fig:BRnew}$b, we can see the enhanced absorption of the transitions $\ket{\pm 3/2}_{\rm g}$ $\leftrightarrow$ $\ket{\pm i/2}_{\rm e}$ (i = 1, 3, 5) due to the increased population in the ground state $\ket{\pm 3/2}_{\rm g}$. The enhanced absorption depth of those antiholes for the transitions $\ket{\pm 3/2}_{\rm g}$ $\leftrightarrow$ $\ket{\pm i/2}_{\rm e}$ (i = 1, 3, 5) can be employed to determine the relative transition strengths. If we burn a narrow hole at $\omega_2$ and sweep the pump light centering at $\omega_1$ in the third step, we can obtain the relative transition probabilities of $\ket{\pm 5/2}_{\rm g}$ $\leftrightarrow$ $\ket{\pm i/2}_{\rm e}$ (i = 1, 3, 5). By repeat such measurements with various spin polarization procedures, we finally obtain the complete relative transition probabilities shown in Tab. 1 in the main text.

\begin{figure}
\includegraphics [width= 0.9 \columnwidth]{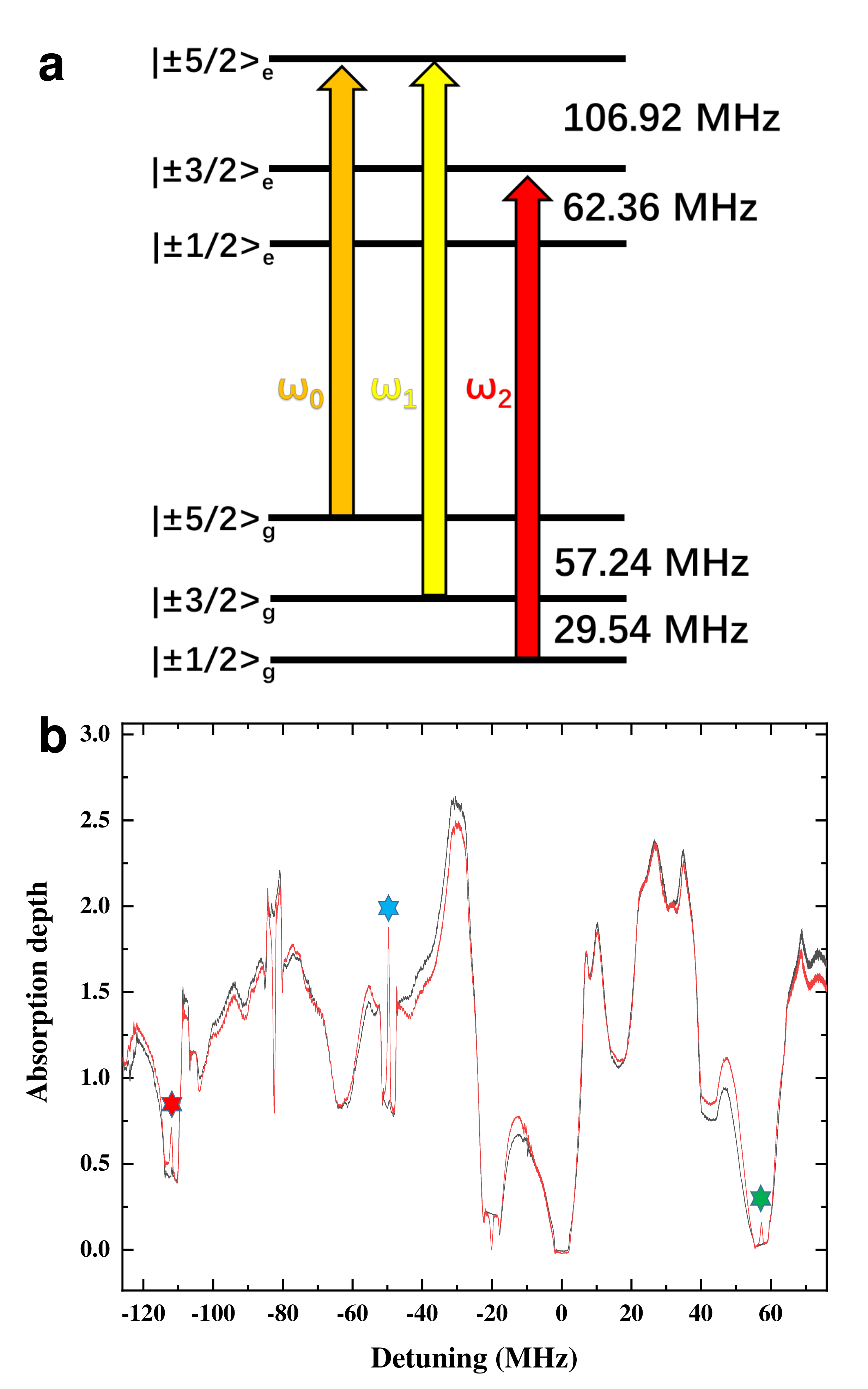}
\caption{\label{fig:BRnew} \textbf{a} An example of the pump light centering at three frequencies to isolate a single class of ions. \textbf{b} An example of the absorption spectrum used for determining the relative transition probabilities. The black line is to the spectrum after the spin polarization. The red line is the final spectrum (see the text for details). Zero detuning corresponds to the frequency $\omega_0$. We can observe three enhanced antiholes, corresponding of the transitions $\ket{\pm 3/2}_{\rm g}$ $\leftrightarrow$ $\ket{\pm i/2}_{\rm e}$ (i = 1, 3, 5) at the frequencies -112.04 MHz (red), -49.68 MHz (blue) and 57.24 MHz (green), respectively. The enhanced absorption depth of these antiholes can be employed to determine the relative transition probabilities for $\ket{\pm 3/2}_{\rm g}$ $\leftrightarrow$ $\ket{\pm i/2}_{\rm e}$ (i = 1, 3, 5). }
\end{figure}

\subsection*{Supplementary Note 3 - \\Experimental details}
\begin{figure}
\includegraphics [width= 0.9 \columnwidth]{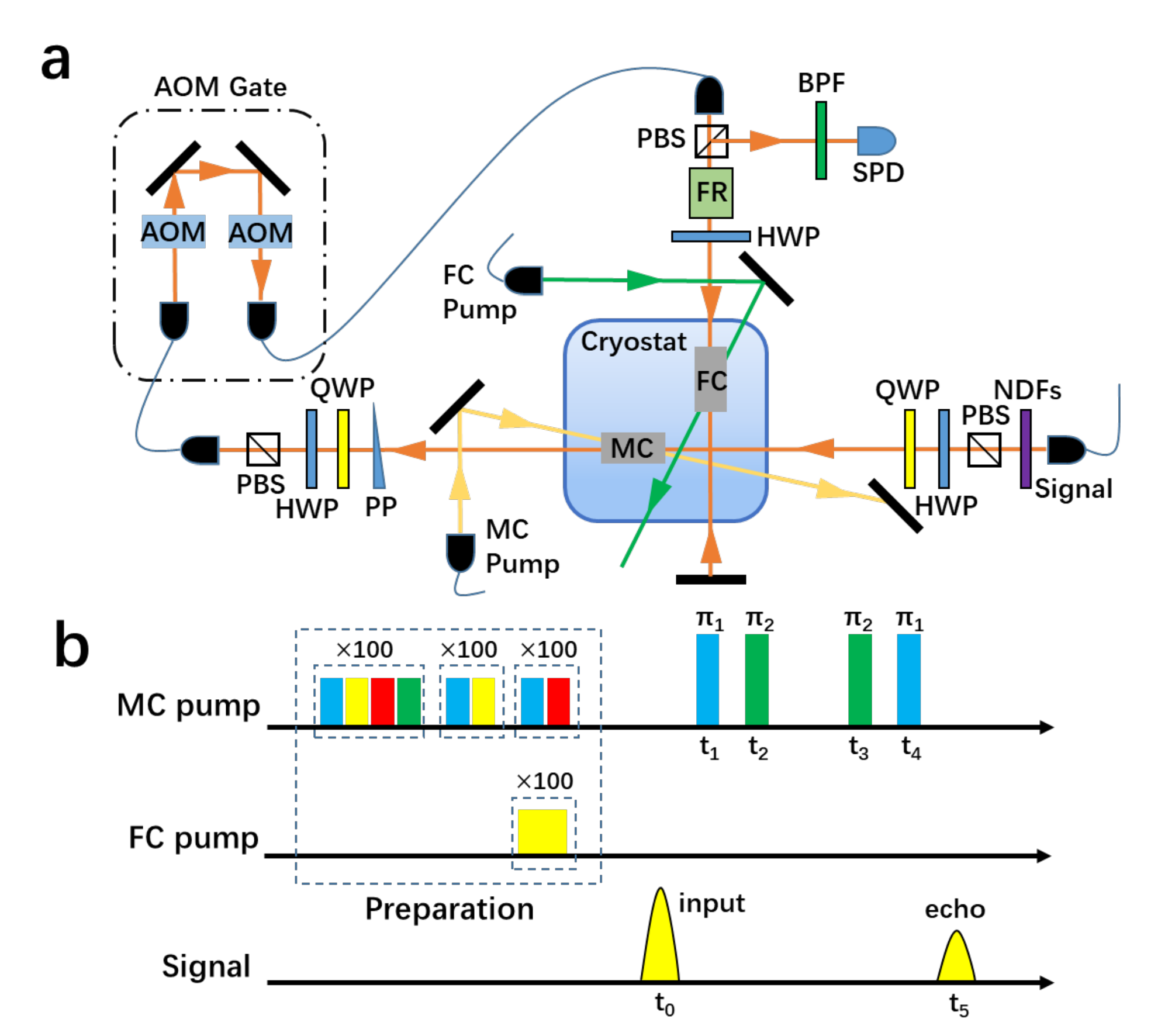}
\caption{\label{fig:Setup} \textbf{a} Schematic of the experimental setup. Several neutral density filters (NDFs) are employed to prepare single-photon-level coherent pulses. The polarization of the signal is prepared by a polarization beam splitter (PBS), a half-wave plate (HWP) and a quarter-wave plate (QWP). After storage in the memory crystal (MC), the polarization of the retrieved echo is then measured using QWP, HWP and PBS. A phase plate (PP) compensates the phase difference along the optical path. An acoustic-optic modulator (AOM) gate consisting of two AOMs serving as a temporal gate. Spectral filtering is realized by two 1-nm band-pass filters (BFP) and a filter crystal (FC) in a double-pass configuration with the help of a PBS and a Faraday rotator (FR). The signal is finally detected by a single photon detector (SPD). \textbf{b} The experimental sequence including a spectral initialization process and the NLPE storage process. }
\end{figure}

The Y$_2$SiO$_5$ crystal belongs to the space group \emph{C$^{6}_{2h}$} and Y$^{3+}$ ions have two different crystallographic sites, which can be occupied by Eu$^{3+}$ ions. Y$_2$SiO$_5$ is a monoclinic and biaxial crystal with two principal axes (D1 and D2) and one optical normal axis (b). As already been observed in Ref. \cite{Eu_2003}, for Eu$^{3+}$ ions at site 2 of Y$_2$SiO$_5$ crystals, the absorption for light with the electric vector parallel to the D1 axis and the b axis is approximately the same for the $^{7}$F$_{0}$ $\leftrightarrow$ $^{5}$D$_{0}$ transition. As a result, direct storage of polarization qubits is in principle possible with site-2 Eu$^{3+}$ ions in a single piece of Y$_2$SiO$_5$ crystal. However, the challenges are two-fold. First, the absorption of site-2 Eu$^{3+}$ ions is much weaker than that of site-1 Eu$^{3+}$ ions in Y$_2$SiO$_5$ crystals \cite{Eu_2003}. The storage protocol must make an efficient use of the original sample absorption. NLPE fulfills this requirement since it requires no spectral tailoring and makes the complete use of the original sample absorption. Second, complete knowledge of the hyperfine energy levels is required for the implementation of a spin-wave quantum memory. Therefore, our experiment starts at the spectroscopic investigation of site-2 Eu$^{3+}$ ions in Y$_2$SiO$_5$ crystals, as detailed in the Supplementary Note 1.

Note that the hyperfine energy differences given by Yano et al. \cite{Yano_92,Yano_91} is consistent with our experimental results while the order of the hyperfine energy levels given in Ref. \cite{Nilsson_2002} is not. The order of the ground state is different from that at site 1, i.e., the larger splitting (57.24 MHz) is above the smaller one (29.54 MHz), and this indicates that the pure quadrupole interaction is stronger than the pseudoquadrupole interaction \cite{MACFARLANE198751}. Based on the known level structure, a single class of ions can be isolated through optical pumping and the relative transition strengths can be determined after spin polarization. The final result is presented in Tab. 1 in the main text and more details are provided in the Supplementalary Note 2.

Gathering the information together, now the medium is ready for the implementation of spin-wave quantum storage. Experimental setup is shown in Supplementary Fig.~\ref{fig:Setup}a. The laser output is divided into three laser beams, denoted as the signal, the memory crystal (MC) pump and the filter crystal (FC) pump beams. The laser source is a frequency-doubled semiconductor laser at 580.6113 nm (516.6968 THz) with a locked linewidth below 1 kHz. All laser beams are modulated by an acoustic-optic modulator in a double-pass configuration. The counter-propagating MC pump beam and the signal beam overlap in the middle of MC with an angle of 25 mrad. The pump beam and the signal beam have a diameter of 180 um and 60 um inside MC, respectively. The maximal power of the pump light for MC is 320 mW. The pump beam for FC has a diameter of 500 um inside FC and the power is 25 $\mu$w. Two crystals are placed in a cryostat with a temperature of approximately 3.4 K.

Since the inhomogeneous broadening (1.5 GHz) of MC is much larger than the hyperfine splittings, a spectral initialization process is required to isolate a well-defined 4-level system for NLPE storage. The spectral initialization process is similar as that presented in Ref. \cite{NLPE_2021}. Pump light at four frequencies, corresponding to the frequencies of the signal, two rephasing pulses and the back pump is employed for preparation of the memory crystal. The transition of the signal should have a strong relative strength for efficient absorption of input photons. The transitions of two rephasing pulses should have the relatively strong strength for high transfer efficiency. According to the data presented in Tab. 1 in the main text, our choice of the four-level system is provided in the Fig. 1a in the main text. Here we choose the transitions $\ket{\pm 1/2}_{\rm g}$ $\leftrightarrow$ $\ket{\pm 1/2}_{\rm e}$ as the signal, $\ket{\pm 3/2}_{\rm g}$ $\leftrightarrow$ $\ket{\pm 1/2}_{\rm e}$, $\ket{\pm 1/2}_{\rm g}$ $\leftrightarrow$ $\ket{\pm 3/2}_{\rm e}$ as two control pulses and $\ket{\pm 5/2}_{\rm g}$ $\leftrightarrow$ $\ket{\pm 3/2}_{\rm e}$ as the back pump light. The center frequencies are f$_{0}$, f$_{1}$, f$_{2}$ and f$_{3}$, respectively.
 
 The experimental sequence is shown in Supplementary Fig.~\ref{fig:Setup}b. In the class cleaning step, the four pump pulses are all chirped by 4 MHz with duration of 1 ms and repeated for 100 times. Only one class of ions (shown in Fig. 1a) is selected in four pits due to the simultaneous resonance of three ground states. In the spin polarization step, two pump pulses centering at f$_0$ and f$_1$ are applied with a chirping bandwidth of 4 MHz. The population is initialized to the state $\ket{\pm 5/2}_{\rm g}$. In the final step, we apply the pump pulse centering at f$_3$ with a bandwidth of 700 kHz to prepare a isolated absorption peak in $\ket{\pm 1/2}_{\rm g}$. In the meantime, the pump light centering at f$_1$ with a chirping bandwidth of 4 MHz is applied to keep the state $\ket{\pm 3/2}_{\rm g}$ empty for spin-wave storage. Meanwhile, the pump light for FC burns a 1-MHz spectral hole in FC, which allows the signal to be completely transmitted and eliminates the noise in frequency domain. The absorption peaks of the MC for the input signal with two polarization states are shown in Supplementary Fig.~\ref{fig:structure}.

After spectral initialization, the input pulse applied at the time t$_0$ establishes an optical coherence between the states $\ket{\pm 1/2}_{\rm g}$ and $\ket{\pm 1/2}_{\rm e}$. For clarity, we denote the rephasing $\pi$ pulses at the frequency f$_i$ as $\pi_i$ (i = 1, 2). The first $\pi_1$ pulse applied at the time t$_1$ converts the optical coherence to the spin coherence between $\ket{\pm 1/2}_{\rm g}$ and $\ket{\pm 3/2}_{\rm g}$. The $\pi_2$ pulse at the time t$_2$ converts the spin coherence to another optical coherence between $\ket{\pm 1/2}_{\rm g}$ and $\ket{\pm 3/2}_{\rm e}$. Due to the phase mismatching \cite{NLPE_2021,ROSE_2011}, the standard 4-level echo \cite{4LPE_2011} expected to emit at the time t$_2$ + (t$_1-t_0$) is silenced. The $\pi_2$ and $\pi_1$ pulses applied at the time t$_3$ and t$_4$ perform the reverse process as that accomplished by the $\pi$ pulses at t$_1$ and t$_2$. The ensemble is double rephased and emits the noiseless echo at the time t$_{5}$ = t$_{4}$ + [(t$_{3}-t_{2}$) $-$ (t$_{1}-t_{0}$)]. The final echo has the same spatial mode and frequency as compared to that of the input signal.

For the specific timing, the NLPE storage process starts with the input at the time t$_0$ = 0. The envelope of the input pulse is a Gaussian function with a full width at half maximum 2.8 $\mu$s. The first $\pi$ pulse is applied at t$_1$ = 6.25 us with a duration of 7.5 us. The second $\pi$ pulse treads on the heels of the first $\pi$ pulse at t$_2$ = 13 us with duration of 6 us. The third $\pi$ pulse incident at t$_3$ = 26 $\mu$s is the same as the second $\pi$ pulse (denoted as $\pi_2$). 
The last $\pi$ pulse incident at t$_4$ = 37.75 us is the same as the first $\pi$ pulse (denoted as $\pi_1$). The echo emits at t$_5$ = 44.5 us. The $\pi_1$ and $\pi_2$ pulse are complex hyperbolic secant pulses \cite{ROSE_2011,NLPE_2021} with the chirping bandwidths of 2.4 MHz and 3.4 MHz, respectively. 

\begin{figure}
\includegraphics [width= 0.9 \columnwidth]{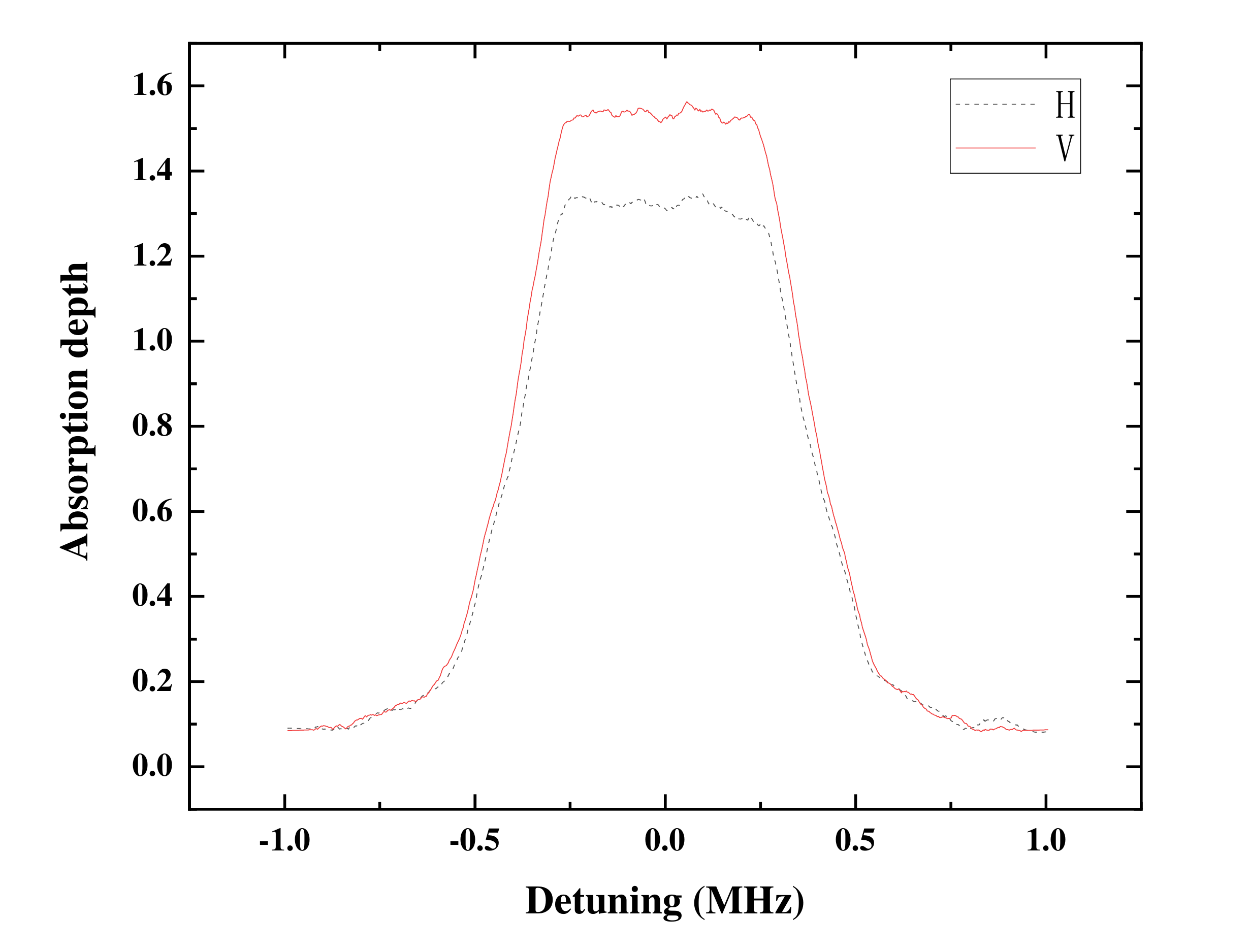}
\caption{\label{fig:structure} The absorption peak of the memory crystal at the center frequency of f$_{0}$. The absorption spectrum is measured with the signal beams of two polarization states $\ket{H}$ (black dotted line) and $\ket{V}$ (red solid line). The absorption depth is 1.53 for $\ket{V}$ and 1.32 for $\ket{H}$, with a bandwidth of approximately 800 kHz.}
\end{figure}

\bibliographystyle{unsrt}

\bibliography{cas-refs}




\end{document}